\documentclass[twocolumn,showpacs,aps,prl,a4paper,amsmath,amssymb]{revtex4}
\usepackage{graphicx}

\begin{document}

\title{Number distributions for fermions and fermionized bosons in periodic potentials}
\author{Michael Budde and Klaus M\o lmer}
\affiliation{ QUANTOP, Danish National Research Foundation Center for Quantum Optics, \\
Department of Physics and Astronomy, University of Aarhus \\
DK-8000 Aarhus C, Denmark}

\begin{abstract}
We compute the spatial population statistics for one-dimensional
fermi-gases and for bose-gases with hard core repulsions in
periodic potentials. We show how the statistics depend on the
atomic density in the ground state of the system, and we present
calculations for the dynamical turn-on of the potential.
\end{abstract}


\maketitle

\section{Introduction}

Since the first achievement of Bose-Einstein condensation (BEC) of
alkali gases in 1995, atomic quantum gas research has continued to
provide important tests of fundamental physics. One very ambitious
current goal for ultra-cold atom research involves cold fermionic
atoms, where association of the fermionic atoms into bose
condensed diatomic molecules has just been reported by several
groups \cite{mol_bec}, and where current efforts aim at the
observation of BCS superfluidity of the gas \cite{ferm_pairs}.
Another goal is to push the study of cold bosons far beyond the
mean field regime, as exemplified already when bosons with a
repulsive interaction are exposed to a periodic potential giving
rise to the transition to a Mott insulator
\cite{mott_theory,Munichdiff}, and in the case of a very dilute
gas in one dimension, where the repulsive interaction effectively
presents a hard core exclusion of overlapping atomic population
and leads to the Tonks-Girardeau regime \cite{tonks_theory}.

The zero temperature state of the Tonks-Girardeau bose-gas
(TG-gas) is described by a wave function which is obtained
mathematically from the Slater determinant describing a spin-less
non-interacting fermi-gas. This has the consequence that these two
systems have many common properties, and in the present paper we
wish to determine one of these properties, namely the fluctuations
in the number of atoms populating a finite region of space. The
population statistics in weakly interacting and in ideal dilute
Bose gases was studied in \cite{scully,condensstat}, which
revealed a pair-correlation mechanism for finite numbers of
non-condensed particles. Measurements of fluctuations represent an
interesting approach to the analysis and understanding of physical
systems which, in analogy with pioneering experiments in quantum
optics, extracts its information from the noise rather than from
mean values in measurements. In other many-body problems, noise
measurements have already been used to demonstrate the fractional
quantum Hall effect, and the Hanbury-Brown and Twiss correlations
in high energy nuclear collision experiments where they may
constitute probes of quark-gluon plasma effects.

We note that a TG-gas recently was produced and observed in
experiments, where a periodic potential was used to alter the
effective mass and hence the importance of the interaction energy
relative to the kinetic energy of the particles
\cite{bloch_tonks}. The analysis in that paper was based on
measurements of the momentum distribution. In this paper we
develop the theory for another observable of this system, namely
the counting noise in real space. We shall also subject our atoms
to periodic potentials and investigate how this alters the
counting statistics.

\section{Spatial population statistics for fermions and TG-bosons}

A pure state of $N$ non-interacting spin-polarized fermionic
particles (i.e., all populating the same internal spin state) is
described by an anti-symmetrized product state (Slater
determinant) of single-particle spatial wave functions

\begin{equation}
\Psi_F(x_1, ..., x_N) \propto {\mbox{det}}\{u_i(x_j,t)\},
\label{slater}
\end{equation}
where the $N$ orthogonal single-particle states $u_i(x,t)$ are all
solutions of the single-particle Schr\"{o}dinger equation of the
atoms. The stationary ground state is thus obtained from the $N$
lowest single-particle energy eigenstates.

A pure state $\Psi_B(x_1, ... ,x_N)$ of $N$ impenetrable bosons
must be symmetric under exchange of the atomic coordinates, and it
must satisfy the condition that it vanishes if any two particle
coordinates coincide: $\Psi_B=0$ if $x_i=x_j$ for any $i\neq j$.
In the limit where the hard-core interaction between the boson has
zero range, a state vector $\Psi_B(x_1, ..., x_N)$ obeying these
constraints is formally obtained by introducing the antisymmetric
function $A(x_1, ... ,x_N)=\Pi_{1\leq i < j \leq N}
{\mbox{sgn}}(x_j-x_i)$, and simply defining,

\begin{equation}
\Psi_B(x_1, ..., x_N) = A(x_1, ..., x_N) \Psi_F(x_1, ..., x_N) .
\label{boseslater}
\end{equation}
This 'fermionization' ansatz is valid both for stationary states
(e.g., the ground state of the system), and for time dependent
problems.

When the state vector is known, the probability that precisely $n$
particles are found in an interval of space $I$, is obtained by an
integration over all configurations contributing precisely to this
possibility, i.e., $n$ coordinates should be within the interval
$I$ and the remaining $N-n$ coordinates should be outside that
interval:

\begin{eqnarray}
p(n) & = & \sum \int_{ x_{i_1}, \ldots x_{i_n}  \in I } \int_{
x_{j_1}, \ldots x_{j_{N-n}}\ \not\in I } dx_1 \ldots dx_N
\nonumber \\ & \times & |\Psi_{F/B}(x_1, \ldots x_N)|^2.
\label{integral}
\end{eqnarray}
The sum extends over all combinations $1\leq i_1 < i_2 ... < i_n
\leq N$, $1\leq j_1 < ... < j_{N-n} \leq N$ with $\forall k,l: i_k
\neq j_l$, and although the different contributions to this sum
all yield the same value, one still has to carry out an
$N$-dimensional integral of  a Slater determinant with $N!$ terms
to obtain the distribution. We have
written the expression (\ref{integral}) here because it shows
explicitly that  non-interacting fermions and
hard-core interacting bosons have identical number
distribution on a given position range. By writing the counting distribution
in terms of atomic creation and annihilation operators, we shall
obtain a simpler expression for the counting statistics.

The treatment below will address the fermionic system and it will
make use of operators with fermionic anti-commutator properties,
but the resulting number distributions on spatial intervals are
applicable to both non-interacting fermions and impenetrable
bosons. Note that the state vectors in momentum space are not
linked as easily as the expressions
(\ref{slater},\ref{boseslater}) in position space:  the free fermi
gas at zero temperature has a flat momentum distribution up to the
fermi momentum, whereas the Tonks gas has a characteristic peaked
distribution around zero momentum. In (\ref{slater}) no
single-particle state is populated by more than a single atom,
whereas the Tonks gas has a single state populated by of the order
$\sqrt{N}$ atoms.

Let $\hat{a}_n$ denote annihilation operators for fermions in a
state with single-particle wave function $u_n(x)$. The connection
between these operators and the position dependent particle
annihilation operator $\hat{\psi}(x)$ is

\begin{equation}
\hat{\psi}(x) = \sum_n u_n(x) \hat{a}_n,\ \ \ \hat{a}_n=\int dx
u_n^*(x)\hat{\psi}(x) \label{operators}
\end{equation}
where we make use of both orthogonality and completeness of the
basis of eigenstates $u_n(x)$.

In the calculations below, we shall assume that the
single-particle states $n=1, \ldots, N$ are all occupied. The
probability to detect a single atom at the location $x_1$ is the
expectation value of the local density operator
$\hat{\psi}^\dagger(x_1)\hat{\psi}(x_1)$, and we can evaluate the
average population of a position interval $I$ as the mean value of
the number operator $\hat{n}_I = \int_{x_1\in I} dx_1
\hat{\psi}^\dagger(x_1) \hat{\psi}(x_1)$,

\begin{equation}
\langle \hat{n}_I\rangle = \sum_{n'=1}^N \int_{x_1\in I} dx_1
|u_{n'}(x_1)|^2.
\end{equation}
Making use of the anti-commutator
$\hat{\psi}(x)\hat{\psi}^\dagger(x')
+\hat{\psi}^\dagger(x')\hat{\psi}(x) = \delta(x-x')$ we can also
address higher order moments of this population,

\begin{eqnarray}
\langle \hat{n}_I^2\rangle & = & \sum_{n'=1}^N \int_{x_1 \in I} dx_1 |u_{n'}(x_1)|^2 \nonumber \\
 & + & \sum_{n'=1}^N\sum_{n''=1}^{n'-1} \int_{x_1 \in I} dx_1 \int_{x_2 \in I} dx_2 \nonumber \\
 &  &  \times |u_{n'}(x_1)u_{n''}(x_2)-u_{n'}(x_2)u_{n''}(x_1)|^2.
\label{nI2}
\end{eqnarray}
If we know the single-particle states occupied by the system, we
can compute the mean population and the variance in the interval
$I$ explicitly. For example, in the lowest energy state of free
atoms, where all the eigenstates are plane waves, we obtain for
the number of fermions/TG-bosons on a 1D interval of length $L$:
$\langle \hat{n}_I \rangle = \rho L$, where $\rho$ is the mean
density of the gas, and we straightforwardly obtain a variance
that scales with the logarithm of the mean occupancy of the
interval. For an infinitely extended system, one can also obtain
this result from the Fermi-Dirac momentum distribution function,
which is the Fourier transform of the spatial density-density
correlation function \cite{castin-priv}, \text{Var}$(n_I) =
(1+\ln(2\pi \rho L)+\gamma)/\pi^2$, with Euler's
$\gamma=0.57721...$ . In quantum optics language, the fermions and
TG-bosons are strongly anti-bunched.

\subsection{Full number distribution}

By use of a technique developed by Levitov \cite{levitov} (for a
simpler derivation of this single aspect of Levitov's results, see
the presentations by Klich \cite{klich} and by Castin
\cite{castin-priv,castin-houches}), it is possible to obtain the full number
distribution $p(n)$ for the number of atoms in the interval $I$,
from which all moments of the distribution are readily calculated.
First, we observe that the operator $\frac{1}{2\pi}\int_0^{2\pi}
d\theta \exp(i\theta(\hat{n}_I-n))$ is the projection operator
onto a subspace where the operator $\hat{n}_I$ has the eigenvalue
$n$, see also \cite{scully}. It is therefore useful to introduce the characteristic
function
\begin{equation}
F(\theta) \equiv \langle \exp(i\theta \hat{n}_I)\rangle,
\label{char}
\end{equation}
from which the number distribution is obtained as a simple Fourier
transform. The expectation value in (\ref{char}) is taken in the
state with $N$ occupied single-particle wave functions, and the
trick is to realize that this state is described by a density
matrix that can formally be written as a thermal state $\sigma =
\exp(-\beta \hat{H})/Z$ where $\beta=1/k_B T$, and where $\hat{H}
= \sum \epsilon_n \hat{a}^{\dagger}_n\hat{a}_n$ is the many-body
(second-quantized) Hamiltonian corresponding to a one-body
Hamiltonian $H$ (for clarity we write operators acting on the
one-particle Hilbert space without hats), which is diagonal in the
basis of states $u_n(x)$. The expectation value in (\ref{char}) is
thus the trace of a product of two exponentials of quadratic forms
in field creation and annihilation operators. In general, if
operators $A,\ B$ and $C$ on the one-particle Hilbert space obey
$\exp(A)\exp(B)=\exp(C)$, then also the relation
$\exp(\hat{A})\exp(\hat{B})=\exp(\hat{C})$ is valid for the
second-quantized expressions $\hat{X}=\sum X_{nm}
\hat{a}^{\dagger}_n \hat{a}_m$, $X=A,\ B,\ C$. Apart from the
factor $1/Z$, our characteristic function is thus the trace of an
operator $\exp(\hat{C})=\exp(\sum_i \lambda_i
\hat{b}^{\dagger}_i\hat{b}_i)$, where $\lambda_i$ are the
eigenvalues of the matrix $C$, and $\hat{b}_i$ are the
corresponding linear combinations of the operators $\hat{a}_n$.
Since the $\hat{b}^{\dagger}_i\hat{b}_i$ operators are Fermi
number operators, their eigenvalues are restricted to the values
zero and unity, and the exponential $\exp(\lambda_i
\hat{b}^{\dagger}_i\hat{b}_i)$ yields the values $1$ and
$\exp({\lambda_i})$. Hence we can write
\begin{eqnarray}
\text{Tr}[e^{\hat{A}}e^{\hat{B}}] & = & \text{Tr} [e^{\hat{C}}] \\
\nonumber & = & \text{Tr} [\exp(\sum_i \lambda_i
\hat{b}^{\dagger}_i \hat{b}_i)] \\ \nonumber & = & \text{Tr}
[\Pi_i \exp(\lambda_i \hat{b}^{\dagger}_i \hat{b}_i)] \\ \nonumber
& =
& \Pi_i(1+e^{\lambda_i}) \\ \nonumber & = & \det[1+e^C] \\
\nonumber & = & \det[1+e^A e^B]  \,
\end{eqnarray}
expressing the trace of exponentials of second-quantized operators
in terms of a determinant involving exponentials of one particle
operators \cite{klich}. We now insert the relevant matrices,

\begin{equation}
F(\theta)=\text{Tr}[e^{i\hat{n}_I \theta} \frac{1}{Z} e^{-\beta
\hat{H}}]=\frac{1}{Z} \det[1+e^{i\theta P_I}e^{-\beta H}],
\label{char2}
\end{equation}
where $P_I = \int_{x \in I} |x\rangle \langle x|$ is a projection
operator ($P_I^2=P_I$), and thus obeys $e^{i\theta
P_I}=1+(e^{i\theta}-1)P_I$. Introducing the population of the
single-particle states, $\pi_n$, the characteristic function then
rewrites
\mbox{$F(\theta)=\det[1+(e^{i\theta}-1)P_I\sum_n\pi_n|n\rangle\langle
n|]$}, and if we represent this operator as a matrix in a discrete
position representation with grid spacing $\Delta x$, so that the
projection $P_I$ effectively extracts only a sub-matrix with $x,x'
\in I$, we get

\begin{equation}
F(\theta)=\det[1+(e^{i\theta}-1) \Delta x \sum_n \pi_n u^*_n(x)
u_n(x')]. \label{char3}
\end{equation}
Knowing the single-particle states, either analytically, by
diagonalization or by propagation of the time-dependent
Schr\"{o}dinger equation, this expression allows us to determine
the population statistics. As stated above, the expression is
valid both for non-interacting fermions and for TG-bosons at zero
temperature, where most of our calculations are performed. In this
case, $\pi_n=1$ for all occupied states. We shall also study how
the statistics depends on temperature. However, because the
Bose-Fermi mapping only holds for $T=0$, these results only apply
to fermions.

Since the determinant is formally a linear combination of products
of $N_x$ matrix elements, where $N_x$ is the number of $x$-values
used to represent the interval $I$, $F(\theta)$ becomes a
polynomial in $e^{i\theta}$ of order $N_x$. This number should be
large enough to resolve the spatial structure of the wave
functions populated, and we see that the Fourier transform assigns
vanishing probability to any number of particles larger than
$N_x$, which is meaningful when we represent the interval by $N_x$
localized basis states. For weakly interacting bosons, their
commutators lead to a generating function which has the
$e^{i\theta}$ dependence in the denominator
\cite{klich,condensstat}, and hence any high occupancy is in
principle possible. In our numerical calculations, we shall of
course check the convergence of our results when larger and larger
$N_x$ values are applied.

For comparison, we note that the number distribution of a pure condensate of
non-interacting bosons is binomial, i.e.
\begin{equation}
p(n) = \frac{N!}{n! (N-n)!} p^n (1-p)^n  \,  \, ,
\end{equation}
where $p = \int_{x \in I} dx |u_0|^2$ is the probability for a
particular particle to be in $I$, as given by the macroscopically
populated single-particle wave function $u_0$ in the condensate.
This result holds irrespective of the external potential, and in
particular, we note that the number distribution for a single
lattice site is independent of the amplitude of the periodic
potential. This ideal bosonic behavior does not apply when the
bosons interact. In the case of repulsive interactions the
condensate exhibits a phase transition to a Mott-insulator with a
specific number of particles pr. lattice site when the lattice
amplitude is raised above a certain value
\cite{mott_theory,Munichdiff}.

\section{Particles in a periodic potential: numerical results}
We shall now use the theory for the spatial statistics to
calculate the number distribution of fermions and TG-bosons in a
single well of a periodic potential.

\begin{figure}
\resizebox{8cm}{!}{\includegraphics{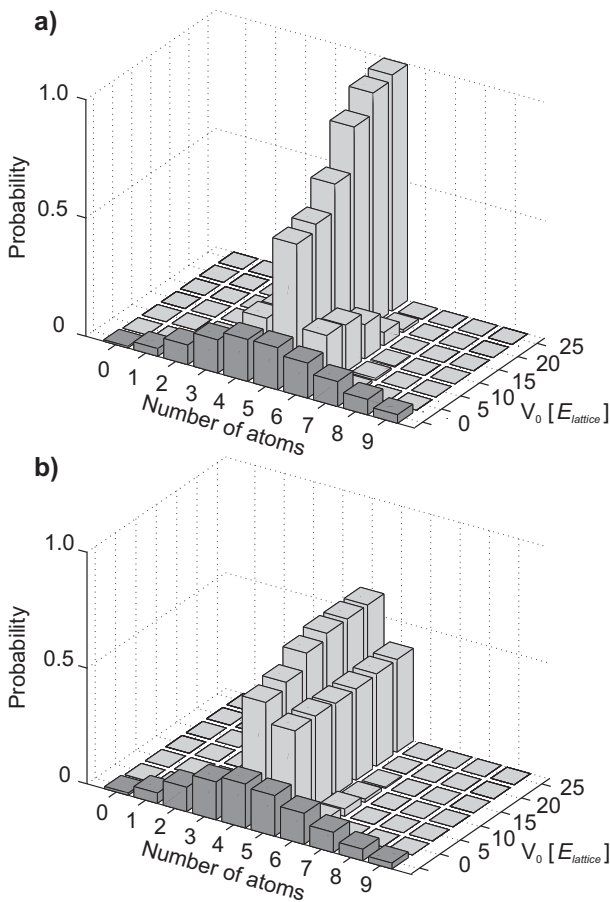}} \caption{Atom
number distributions for fermions and TG bosons in a single
potential well as a function of the lattice amplitude $V_0$. a)
mean occupation $\bar{n}=5$ and b) mean occupation $\bar{n}=4.4$.
The distributions were calculated assuming $T=0$ and that the
periodic potential was turned on adiabatically. The amplitude is
given in units of the lattice energy $E_{lattice}$ defined in the
text. For comparison, the distributions for non-interacting bosons
(dark grey bars) are shown at the same mean occupancies.}
\label{fig1}
\end{figure}

\subsection{Statistics of the ground state}
We first investigate the ground state properties of the system,
i.e. we assume that the $N$ particles occupy the $N$
single-particle states with lowest energy. To find the wave
functions $u_n(x)$ of these states required to obtain the number
statistics by Fourier transformation of (\ref{char3}), we perform
a band structure calculation to solve the single-particle
Schr\"{o}dinger equation
\begin{equation}
-\frac{\hbar^2}{2 m}\frac{d^2u_{n,q}}{dx^2} + \frac{1}{2} V_0 (1 +
\cos[\kappa x])u_{n,q} = \hbar\omega_{n,q} u_{n,q}  \,  \, .
\label{stat_schrodinger}
\end{equation}
Here, $m$ is the atomic mass, and $\kappa = 4\pi/\lambda$, where
$\lambda$ is the wavelength of the optical lattice beams used to
create the periodic potential. $V_0$ is the amplitude of the
periodic potential, which is proportional to the intensity of the
lattice beams and therefore adjustable experimentally. The natural
unit for the amplitude is the lattice energy $E_{lattice} = 2
h^2/(m \lambda^2)$, which differs from the recoil energy
associated with the optical lattice beams by a factor of four. Due
to the periodicity of the Hamiltonian, the wave functions take the
Bloch form $u_{n,q}(x)= \exp(i q x) \phi_{n,q}(x)$, where the
integer $n$ is the band index and $q$ is the crystal momentum,
which under the assumption of periodic boundary conditions over
$n_w$ periods, can take $n_w$ discrete values usually taken to be
in the first Brillouin zone: $q = \frac{4 \pi}{\lambda}\frac{n_q -
n_w/2}{n_w}$ , $n_q \in \{1,2, \ldots ,n_w\}$
\cite{ascroft_mermin}. The function $\phi_{n,q}(x)$ is periodic
with a spatial period $\lambda/2$. We expand it on a plane-wave
basis set $\phi_{n,q}(x) = \sum_{n_b=1}^{2 n_b^{max} +1}
c_{n,n_b}(q) \exp[i \frac{4 \pi}{\lambda}x(n_b - n_b^{max} - 1)]$,
and solve the Schr\"{o}dinger equation by finding the coefficients
$c_{n,n_b}(q)$ of the eigenvectors of the Hamiltonian in this
basis. Once the eigenvectors are known for sufficiently many $q$,
and as a function of $V_0$, the effect of the periodic potential
on the number statistics can be calculated by insertion of the
wave functions into (\ref{char3}) and by calculating the Fourier
transform of $F(\theta)$.

\begin{figure}
\resizebox{8cm}{!}{\includegraphics{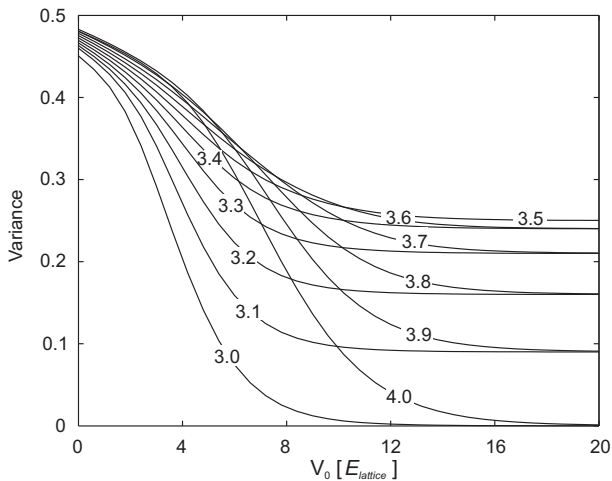}}
\caption{Variance of the number distribution for fermions and
TG-bosons at a single lattice site as a function of lattice
amplitude $V_0$ for various mean occupancies
$\bar{n}=n_{int}+\epsilon$. The variance converges towards
$\epsilon (1 - \epsilon)$ at large amplitudes. The amplitude
required for convergence increases with $\bar{n}$. The
calculations assume that the periodic potential is raised
adiabatically and that $T=0$.} \label{fig2}
\end{figure}

Figure~\ref{fig1} shows the number distribution for fermions and
TG-bosons at a lattice site as a function of $V_0$ for two
selected mean occupancies $\bar{n} = 5$ and $\bar{n} = 4.4$. Also
shown in the figure are the distributions of non-interacting
bosons at the same mean occupancy (dark grey bars). For free
particles ($V_0 = 0$), the number distribution of fermions and
TG-bosons is already much narrower than that of non-interacting
bosons. Moreover, the distribution of TG-bosons and
non-interacting fermions changes with $V_0$, in contrast to
non-interacting bosons. In the case of commensurate filling
($\bar{n}$ integer), the distribution becomes $P(n) =
\delta_{n,\bar{n}}$ for large $V_0$, and consequently the variance
of the atom number goes to zero. This is exactly what one would
expect for fermions because the Pauli principle makes it
energetically favorable to distribute the particles evenly over
the lattice sites. When the filling is incommensurate ($\bar{n} =
n_{int} + \epsilon$, where $n_{int}$ is the integer-part of
$\bar{n}$), the distribution becomes bimodal at large $V_0$, with
$P(n_{int})=(1-\epsilon)$ and $P(n_{int}+1)=\epsilon$. Hence, the
variance converges towards the value $\text{Var}_I = \epsilon (1 -
\epsilon)$, which is symmetric around $\epsilon=0.5$ and
independent of $n_{int}$. The dependence of $\text{Var}_I$ on
$V_0$ is shown in Fig.~\ref{fig2} for a selection of mean
occupancies in the range $ 3 \leq \bar{n} \leq 4$. In addition to
the symmetry of the variance around $\epsilon=0.5$ mentioned
above, the figure shows that the lattice amplitude required for
the probability distribution to converge increases with the mean
occupation. To address this issue quantitatively, we calculated
the amplitude $V_0^{\text{var=1\%}}$ required for the variance to
be reduced to 1\% of the variance for free particles as a function
of $\bar{n}$ for integer $\bar{n}$. Our findings are listed in
Table~\ref{table1}, which also compares $V_0^{\text{var=1\%}}$ to
the mean energy of the highest occupied band $\langle E
\rangle_{band}$ at a lattice amplitude $V_0^{\text{var=1\%}}$. As
one might expect, the lattice amplitude has to be slightly larger
than the mean energy of the highest occupied band in order to
reduce the variance to the few-\% level.

\begin{table}
\begin{tabular}{c c c c c c c c c}
$\bar{n}$           & 1    & 2    & 3   & 4    & 5    & 6    & 7    & 8     \\
$V_0^{\text{var=1\%}}$ [$E_{lattice}$] & 2.5  & 5.9  & 10.7 & 16.8 & 24.1 & 32.8 & 42.7 & 53.8   \\
$\langle E \rangle_{\text{band}}$ [$E_{lattice}$]& 0.7 & 3.3  &
7.2 & 12.5 & 19.1 & 26.9 & 36.1 & 46.5
\end{tabular}
\caption{The row $V_0^{\text{var=1\%}}$ lists the lattice
amplitude required to reduce the variance of the atom number
distribution at a lattice site to 1\% of the variance for free
particles as a function of the mean occupancy $\bar{n}$. The row
$\langle E \rangle_{\text{band}}$ contains the mean energy of the
highest occupied band at a lattice amplitude of
$V_0^{\text{var=1\%}}$. The energy unit is $E_{lattice}$.}
\label{table1}
\end{table}

\subsection{Effect of non-adiabatic turn-on}

So far we have only considered the ground state of the system,
i.e. we have assumed that the periodic potential is turned on
\emph{adiabatically}, so that only the $N$ single-particle states
with lowest energy are occupied at all times. In reality, the
potential will be turned on using a specific ramp $V_0(t)$, and if
the ramp is too fast, atoms will be driven into initially
unoccupied bands, which will lead to an increase in the atom
number fluctuations at a lattice site. We shall now analyze how
the lattice turn-on affects the state and the spatial statistics
of the many-body system. To this end, we calculate the
time-evolution of the state of the system for various ramps by
solving the time-dependent Schr\"{o}dinger equation for the
single-particle states.

Our procedure is as follows: At each time $t$, we expand the
single-particle states $\phi^{\text{dyn}}_{n,q}(t)$ on the Bloch
eigenfunctions $\phi^{(0)}_{n',q'}[V_0(t)]$ at the amplitude
$V_0[t]$ set by the ramp at that particular time:
\begin{equation}
\phi^{\text{dyn}}_{n,q}(t)=\sum_{n',q'} b_{n,q;n',q'}
\phi^{(0)}_{n',q'} ~e^{-i \int_{0}^{t} \omega_{n',q'}(t') dt'}  \,
\, ,
\end{equation}
where $\hbar \omega_{n',q'}(t')$ is the energy of the state
$\phi^{(0)}_{n',q'}[V_0(t')]$.

This expansion is inserted into the time-dependent Schr\"{o}dinger
equation which results in a set of first-order differential
equations for the time dependence of the coefficients
$b_{n,q;n',q'}$.


To solve these differential equations, we need to know the matrix
elements $\langle \phi^{(0)}_{n',q'} | \frac{d}{dt}
\phi^{(0)}_{n'',q''} \rangle$ and the band structure
$\omega_{n',q'}$ as a function of $V_0(t)$. We therefore start our
calculations by solving the static Schr\"{o}dinger equation
(\ref{stat_schrodinger}) for a sufficiently large number of
$V_0$'s ranging from zero to the final $V_0$ of the ramp. We then
calculate the matrix elements:
\begin{equation}
\langle \phi^{(0)}_{n',q'} | \frac{d}{dt} \phi^{(0)}_{n'',q''}
\rangle = \frac{dV_0}{dt} \sum_{n_b=1}^{2 n_b^{max}+1}
c_{n',n_b}^{*}(q') \frac{d c_{n'',n_b}(q'')}{dV_0} \delta_{q',q''}
\, \, . \label{matrixelements}
\end{equation}
Here, $c_{n',n_b}(q')$ are the plane wave expansion coefficients
introduced previously. Note that the matrix elements are diagonal
in the crystal momentum quantum number and, consequently, the
dynamics can be solved for each $q$ independently. Once the
matrix-elements and energies of the differential equations have
been calculated, the coefficients $b_{n,q;n',q'}$ are propagated
in time. We assume that the system is in the ground state
initially, i.e. \mbox{$b_{n,q;n',q'}(t=0)=\delta_{n,n'}
\delta_{q,q'}$} for $\{n,q\}$ corresponding to the $N$ states with
lowest energy and \mbox{$b_{n,q;n',q'}(t=0)=0$} for all other
states.

To quantify the degree of excitation caused by the turn-on, the
time-evolved single-particle states are used to calculate the
population of those Bloch states, which are not populated
initially. To make this number independent of the number of
periods $n_w$ for the periodic boundary condition, we shall
discuss the degree of excitation in terms of the {\it number of
excited atoms pr. period} of the periodic potential
$N_{\text{exc}}$. To investigate how the excitations affect the
atom number statistics, we also insert the time-evolved
single-particle states into (\ref{char3}). There are infinitely
many ways to turn on the optical lattice. We shall limit our
discussion to two cases: linear and exponential ramps. The linear
ramps start at $V_0 = 0$ and increase linearly to the final value
$V_{0,final} = 60 E_{lattice}$ over a time interval of length
$t_{\text{ramp}}$. In the experimentally relevant case of
$^{40}\text{K}$ fermions and an optical lattice wavelength of
$\lambda=850$~nm, we shall study ramp times up to 80
$\hbar/E_{lattice} \sim 461~\mu$s.  Our exponential ramps have the
time-dependence $V_0 (t)=V_{0,final} [\exp(5 t/t_{\text{ramp}}) -
1 ]/[\exp(5)-1]$, $t \in [0,t_{\text{ramp}}$]. This ramp is very
similar to the ones used in the recent experimental realization of
a TG-gas \cite{bloch_tonks}.

\begin{figure}
\resizebox{8cm}{!}{\includegraphics{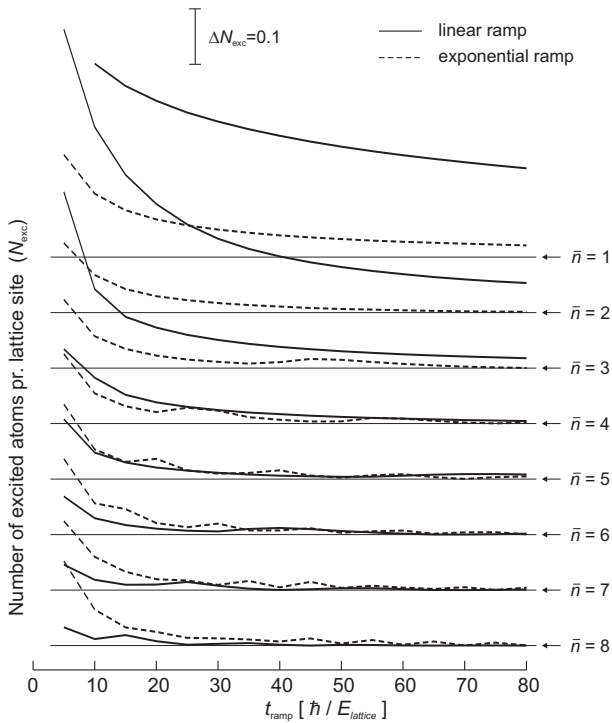}} \caption{Number of
excited atoms pr. lattice site as a function of the ramp time and
of the mean number of atoms pr lattice site (for commensurate
filling). The solid (dashed) curves are the results for linear
(exponential) time dependencies. For clarity, the curves for
different $\bar{n}$ are offset vertically with a spacing of
$\Delta N_{\text{exc}}=0.1$. Thus, the spacing between horizontal
lines defines the scale for the vertical axis. The duration of the
ramp is given in units of $\hbar/E_{lattice}$, where $E_{lattice}$
is the lattice energy defined previously. For $^{40}$K and an
optical lattice wavelength of $\lambda = 850$~nm,
$\hbar/E_{lattice} = 5.8 \mu\text{s}$ } \label{fig3}
\end{figure}

To study the cross-over from nonadiabatic to adiabatic turn-on and
the effect on the atom number statistics for a single lattice
site, we calculated the time evolution for linear and exponential
ramps with durations $5 \hbar/E_{lattice} \leq t_{\text{ramp}}
\leq 80 \hbar/E_{lattice}$ and for various mean occupancies
$\bar{n}$. The final amplitude was constant $V_{0,final}=60
E_{lattice}$ for all calculations presented here. Fig.~\ref{fig3}
shows the number of excited atoms pr lattice site
($N_{\text{exc}}$) as a function of $t_{\text{ramp}}$ for integer
mean occupancies in the range $ 1 \leq \bar{n} \leq 8$. The solid
curves are the results for linear ramps and the dashed curves
represent exponential ramps.  The curves corresponding to
different $\bar{n}$ are offset vertically from each other by
$\Delta N_{\text{exc}}=0.1$, and for each $\bar{n}$, the thin
horizontal solid line corresponds to $N_{\text{exc}}=0$. Thus, the
spacing between the horizontal lines defines the vertical scale of
the figure. Starting in the lower right corner of the figure, we
see that for large mean occupancies $\bar{n} \in {7,8}$, and ramp
times larger than $\sim40 \hbar/E_{lattice}$, the number of
excited atoms pr lattice site is at or below the few-\% level.
Thus, the turn-on is essentially adiabatic. Moreover, linear ramps
lead to less excitation than the corresponding exponential ramps.
This also holds for faster ramps, where the degree of excitation
increases. When the mean number of atoms pr site is decreased to
$4 \leq \bar{n} \leq 6$, the excitations caused by the two types
of ramps are quite similar. For small mean occupancies, $\bar{n}
\leq 3$, exponential ramps are superior because they rise more
slowly at early times where the ramp surpasses the energy of the
occupied bands.

Our adiabatic calculations show that (cf. Fig.~\ref{fig2})
partially occupied bands (incommensurate filling) lead to
fluctuations in the number of atoms pr lattice site even at large
$V_0$. Non-adiabatic turn-on of the periodic potential transfers a
fraction of the atoms from the highest occupied to the lowest
unoccupied band, and therefore results in (at least) two partially
filled bands, which leads to fluctuations in the site occupancy
also in the case of  commensurate filling. Using the time-evolved
single-particle states and our theory for the spatial statistics,
we find that the probability distribution at large $V_0$ for
integer $\bar{n}$ broadens compared to the adiabatic monomodal
distribution $p(n)=\delta_{n,\bar{n}}$. Except for (a) very fast
exponential ramps ($t_{\text{ramp}} \leq 10 \hbar/E_{lattice}$)
and (b) ramp durations of $t_{\text{ramp}} \leq 15
\hbar/E_{lattice}$ and small mean occupancies $\bar{n} \leq 2 $
for linear ramps, the distribution is to a good approximation
given by $\{p(\bar{n}-1)=\eta,p(\bar{n})= 1 - 2\eta,
p(\bar{n}+1)=\eta \}$, where $\eta = N_{\text{exc}}$. This
distribution has a variance of $2 \eta$, and the variance can
therefore be obtained from Fig.~\ref{fig3} simply by scaling
$N_{\text{exc}}$ by a factor of two.

\subsection{Effect of non-zero temperature}

All the results presented so far correspond to zero temperature.
In this last section we shall discuss how non-zero temperature
affects the spatial statistics of fermions in periodic potentials.
The bose-fermi mapping (\ref{boseslater}), which is the reason
that the spatial statistics of TG-bosons is identical to that of
fermions, does not apply at non-zero temperature. Our theory and
the results presented in this section does therefore not apply to
TG-bosons.

It is straightforward to generalize our numerical simulations to
non-zero temperature. The only difference lies in the occupation
probabilities of the states $\pi_n$ in (\ref{char3}), which for
non-zero $T$ are given by the Fermi-Dirac distribution:
\mbox{$\pi_{n,q} = (\exp[ \beta (\hbar \omega_{n,q} - \mu)]  + 1
)^{-1}$}, where $\mu$ is the chemical potential defined by the
constraint that the sum of $\pi_{n,q}$ over all states equals the
appropriate total number of particles. We assume that the system
is in thermal equilibrium with a temperature $T$ and at the
amplitude $V_0$, i.e. our results correspond to the case were the
system has equilibrated at $V_0$. Alternatively one might have
assumed that the system equilibrates so slowly that it retains the
distribution $\pi_{n,q}$ corresponding to free particles
($V_0=0$), and then calculate the statistics using these
occupation probabilities. We shall only consider the first case
here.

\begin{figure}
\resizebox{8cm}{!}{\includegraphics{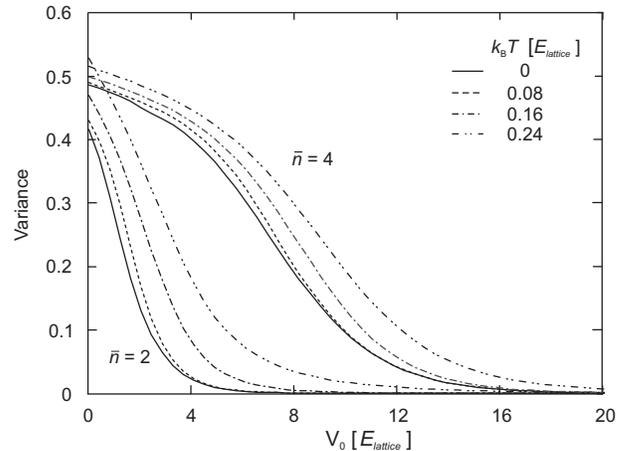}} \caption{Variance
of the fermion number distribution as a function of $V_0$ for
$\bar{n}=\{2,4\}$ and temperatures $T=\{0,0.08, 0.16, 0.24\}
E_{lattice}/k_{\text{B}}$. } \label{fig4}
\end{figure}

Assuming that the periodic potential has been raised
adiabatically, there are two contributions to the fluctuations in
the number of particles at a lattice site. In addition to the
fluctuations due to the fermionic nature of the particles, which
is present at $T=0$ and was discussed above, thermal excitation
lead also lead to fluctuations. This can be seen in
Fig.~\ref{fig4}, which shows the variance as a function of $V_0$
for two mean occupancies ($\bar{n}=\{2,4\}$) and as a function of
temperature. The temperature is given in units of
$E_{lattice}/k_{\text{B}}$. In the case of $^{40}\text{K}$ and
$\lambda = 850$~nm, the four temperatures are 0~nK, 120~nK,
240~nK, and 360~nK. According to the figure, the variance
increases with temperature for a given $V_0$ due to thermal
excitations. For small $V_0$, the increase in variance due to
thermal excitations decreases when particles are added to the
system. The reason is that the mean spacing between bands
increases with increasing band index. Since the Fermi-Dirac
distribution changes from unity to zero over an energy range
$k_{\text{B}}T$, the increase in energy level spacing leads to
less thermal excitation. Fig.~\ref{fig4} also shows that the
variance, for the case of commensurate filling, goes to zero when
the temperature is non-zero. The amplitude $V_0$ required to
obtain a certain variance increases with $T$. The reason for this
behavior is also easily understood. At large $V_0$, the particles
are localized near the periodic potential minima, and the
potential can be expanded to second order around the minima. In
this regime the bands a equally spaced with an energy spacing
proportional to $\sqrt{V_0}$. Thus, the energy spacing between
bands increases with $V_0$ and, consequently, the thermal
fluctuations vanish for large $V_0$.

The thermal contribution to the atom number distribution can be
studied separately by choosing $V_0$ such that the variance at
$T=0$ is much less than the variance at the temperature of
interest. In this regime, the atom number distribution for
commensurate filling is $\{p(\bar{n}-1)=\eta,p(\bar{n})= 1 -
2\eta, p(\bar{n}+1)=\eta \}$, like in the case of excitations
caused by non-adiabatic turn-on. In the thermal case, $\eta$ is
equal to $\sum_{n,q} \pi_{n,q}/n_w$, where the sum extends over
all the states not occupied at $T=0$.

\section{Conclusion}

We have presented a theory for the spatial statistics of fermions
and Tonks-Girardeau bosons in one dimension, and applied the
theory to the case of fermions and TG-bosons in periodic
potentials. The number distribution of fermions and TG-bosons
occupying a single lattice site is significantly narrower than
that of non-interacting bosons. Moreover, the distribution depends
on the amplitude of the periodic potential and becomes bimodal
(monomodal for commensurate filling) at amplitudes on the order of
the energy of the highest occupied band. This behavior is
completely different from that of non-interacting bosons, which is
independent of the potential depth.

We also studied the dynamics associated with the turn-on of the
periodic potential and the effect of excitations on the number
distribution of a lattice site. Non-adiabatic turn-on of the
potential leads to partially filled bands and fluctuations in the
site occupancy even in the case of commensurate filling. For
commensurate filling and most ramps of practical interest, the
probability of having one atom more or less than the mean is equal
to the average number of excited atoms pr lattice site. The
variance is well approximated by twice this number.

Finally, we studied how thermal excitations affect the statistics
of fermions in periodic potentials. We found that the fluctuations
in the number of atoms at a lattice site increase with
temperature. The variance of the number distribution converges
towards the variance at $T=0$ when the potential depth is
increased, but the potential depth required to obtain a certain
variance increases with temperature. For commensurate filling and
lattice depths large enough that the fluctuations at $T=0$ are
vanishingly small, thermal excitations lead to a symmetric atom
number distribution. Like in the case of non-adiabatic turn-on,
the probability of having one atom more or less than the mean is
to a good approximation equal to the mean number of excited atoms
pr lattice site.

\begin{acknowledgments}
Yvan Castin is acknowledged for stimulating discussions and for
referring us to the work on the general population statistics.

\end{acknowledgments}


\begin{thebibliography}{99}
\bibitem{banddispersion}At the lattice amplitudes required to
reduce the variance to a few percent of the variance at $V_0=0$
the bands are essentially flat, i.e. the energies $ \hbar
\omega_{n,q}$ of a given band are nearly equal.
\end{thebibliography}
\end{document}